# Light-matter interactions in chip-integrated niobium nano-circuit arrays at optical fibre communication frequencies

Kaveh Delfanazari [1,*] & Otto L. Muskens [2]

1. James Watt School of Engineering, University of Glasgow, Glasgow G12 8QQ,UK

2. Physics and Astronomy, University of Southampton, Southampton SO17 1BJ, UK

*corresponding author: kaveh.delfanazari@glasgow.ac.uk

Dated:01062021

**The interplay between electronic properties and optical response enables the realization of novel types of materials with tunable response. Superconductors are well known to exhibit profound changes in the electronic structure related to formation of Cooper pairs, yet their influence on the electromagnetic response in the optical regime has remained largely unstudied. Photonics metamaterials offer new opportunities to enhance the light-matter interaction, boosting the influence of subtle effects on the optical response. Combination of photonics metamaterials and superconducting quantum circuits will have then potential to advance the quantum computing and quantum communication technologies. Here, we introduce subwavelength photonic nano-grating circuit arrays on the facet of niobium thin films to enhance light matter interaction at fibre optic communication frequencies. We find that optical resonance shifts to longer wavelengths with increasing nano-grating circuit periodicity, indicating a clear modulation of optical light with geometrical parameters of the device. Next to the prominent subwavelength resonance, we find a second feature consisting of adjacent dip and peak appears at slightly shorter wavelengths around the diffraction condition $P_y = \lambda$, corresponding to the Wood's and Rayleigh's anomalies of the 1st order grating diffraction. The observed tunable plasmonic photo-response in such compact and integrated nano-circuitry enables new types of metamaterial and plasmonics based modulators, sensors and bolometer devices.**

In recent years, the application of plasmonics and metamaterials has enabled new research in photonic devices, communications, quantum information and biotechnologies. Traditional plasmonic metals such as gold and silver have been predominantly used because of the low rates of free electron scattering which results on high conductivity in the RF range and high reflectivity in the optical range [1]. There is however an increased interest in using alternative materials that combine a free-carrier optical response with other complex physical characteristics, such as magneto-optic activity, phase-change response, correlated electron systems, and superconductors with strongly tunable superfluid plasma frequency [2-18].

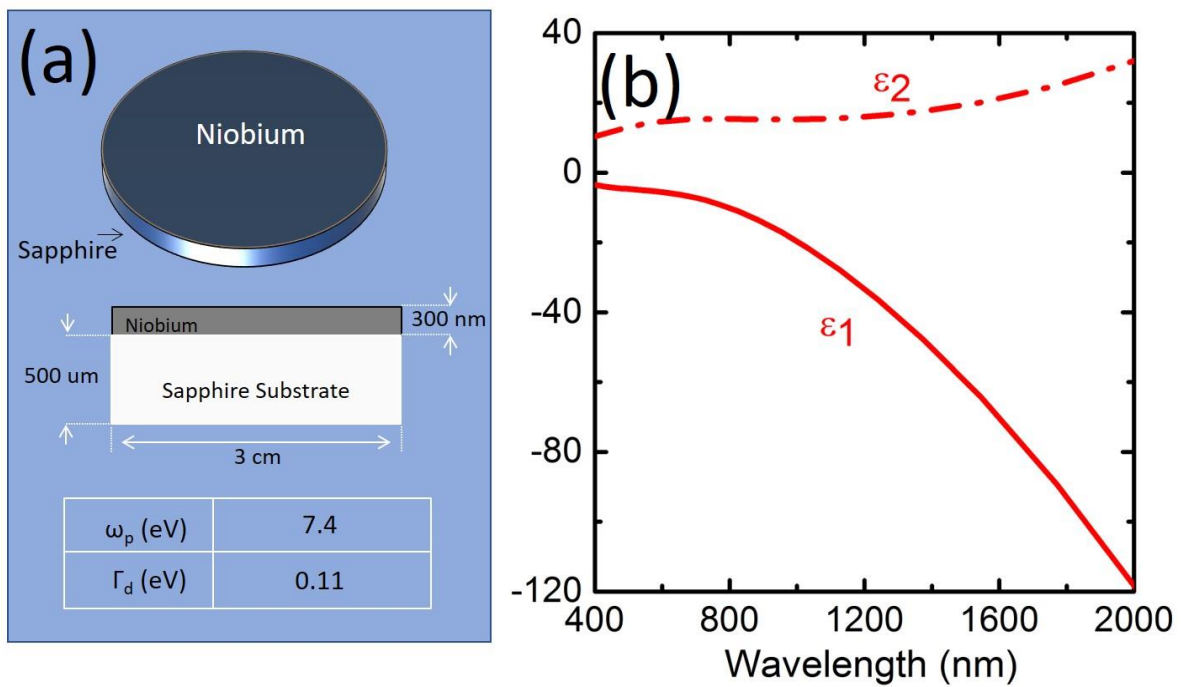

**Figure 1. Optical and plasmonic properties of the niobium (Nb) in the optical telecommunication frequencies:** (a) Artistic impression of the Nb chip with related information. (b) Complex permittivity of 300 nm thick unstructured Nb film deposited on the sapphire substrate measured by spectroscopic ellipsometry at room temperature. Negative real part of permittivity of Nb is clearly seen for whole ultraviolet, visible and near infrared frequency ranges.

Superconductors with zero dc and low transient frequency losses are plasmonic materials [2,5,6,16-18]. Their unique electromagnetic response governed by superconducting electrons (Cooper pairs) is strongly sensitive to the external perturbations such as magnetic fields

[19,20], current [21], voltage [6], temperature [6,7, 9, 22-29] and light [4]. There have been significant studies on highly tunable superconducting metamaterials [9,27] and interesting physical phenomena including Fano resonances [28], electromagnetically induced transparency [6,29], quantum interference effect [19,20,30] and plasmonic response [5,6,16,22,25,26,31,32] have been reported. Most of the theoretical and experimental studies have been performed for the RF, microwave, millimetre and terahertz (THz) wave regime based on common assumption that light quanta with energies above few *me*V which (depends on the superconductor material type) is greater than the superconducting energy gap breaks cooper pairs, supresses the superconductivity and therefore turns superconducting state to normal. As a result, the optical properties of superconducting materials including bulk and thin film, superconducting circuits, and superconducting nanostructures, especially at optical fibre telecommunication are remarkably unexplored. This is an important issue in quantum technology where superconducting quantum circuits have been found as leading technology for quantum computing and quantum communication [3, 33].

The integration, scale-up, and multiplexing arrays of quantum devices in a single chip are the main challenges of superconducting based quantum technology. To overcome the heat load associated with the increasing number of quantum processors in cryogenic temperatures photonic links has been recently proposed [3]. The platform utilizes optical fibres, with their low thermal conductivity, to connect superconducting quantum hardware nodes with optical fibre telecommunication system. Therefore, it is vital to investigate the optical properties of the superconducting materials and the devices interfacing superconducting circuits and optical fibres.

The electrodynamics of niobium (Nb) and aluminium (Al) have been studied both experimentally and theoretically, placing them in the category of metallic superconductors [17,18]. Metallic superconductors are well studied superconducting materials especially in quantum technology. They form the building block for cryogenic communication circuits [6] and superconducting quantum chips [34], etc. Nb based devices when formed as nanostructures show interesting optoelectronic properties under visible and near infrared light illumination [35]. Al-based highly-doped InAs nanowire Josephson junctions have been found to be robust to illumination by optical photons [36].

Here, photonics metamaterials are used to enhance the light-matter interaction. We fabricated subwavelength photonic nano-grating circuit arrays on the facet of niobium thin films. We study the light-matter interaction at fibre optic communication frequencies. A Nb thin film of 300 nm thickness was deposited onto a sapphire substrate of 30 mm diameter and 0.5 mm thickness. The schematic view of the Nb chip with related information is shown in Fig. 1 (a).

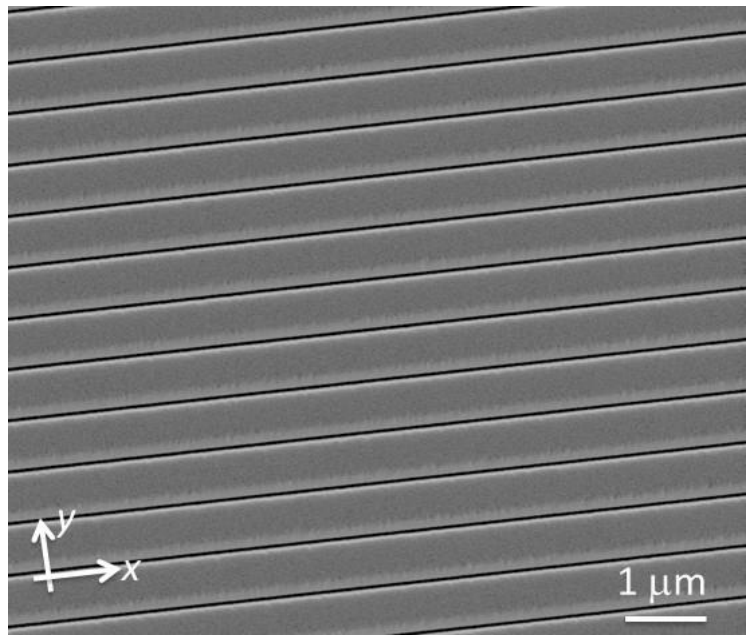

**Figure 2:** The scanning electron microscope (SEM) image (top view) of the Nb optical nano-circuit array with width $w$= 200 nm, length $L$= 100 mm, Period $P_y$= 900 nm fabricated by using focused ion beam (FIB) milling.

Optical reflectivity experiments at optical fibre telecommunication frequencies were performed with the Nb sample mounted inside an optical setup and continuous waves (cw) white light with relatively low intensity was focused to an approximately 30 μm diameter spot on the sample's surface and reflection of the light and its corresponding extinction spectra $A(\lambda)$

= 1- R(λ) at normal incidence was detected by using a spectrophotometer, for two incident polarizations of the electric field parallel (*x*-polarized) and perpendicular (*y*-polarized) to the nano-circuits. Figure 1 (b) shows the dielectric function of the unstructured Nb film at room temperature, measured by variable-angle spectroscopic ellipsometry in the range 200 nm - 2000 nm.

A negative permittivity of Nb is observed over the whole UV-VIS-NIR range. The ratio of the real and imaginary parts $\varepsilon_1/\varepsilon_2$ represents the role of losses and exceeds 10 for gold and silver over the near-infrared range. Nb shows a $\varepsilon_1/\varepsilon_2$ ratio exceeding 2 only for wavelengths above 1200 nm, indicating that it will support plasmonic modes which are around five times stronger damped than for noble metals.

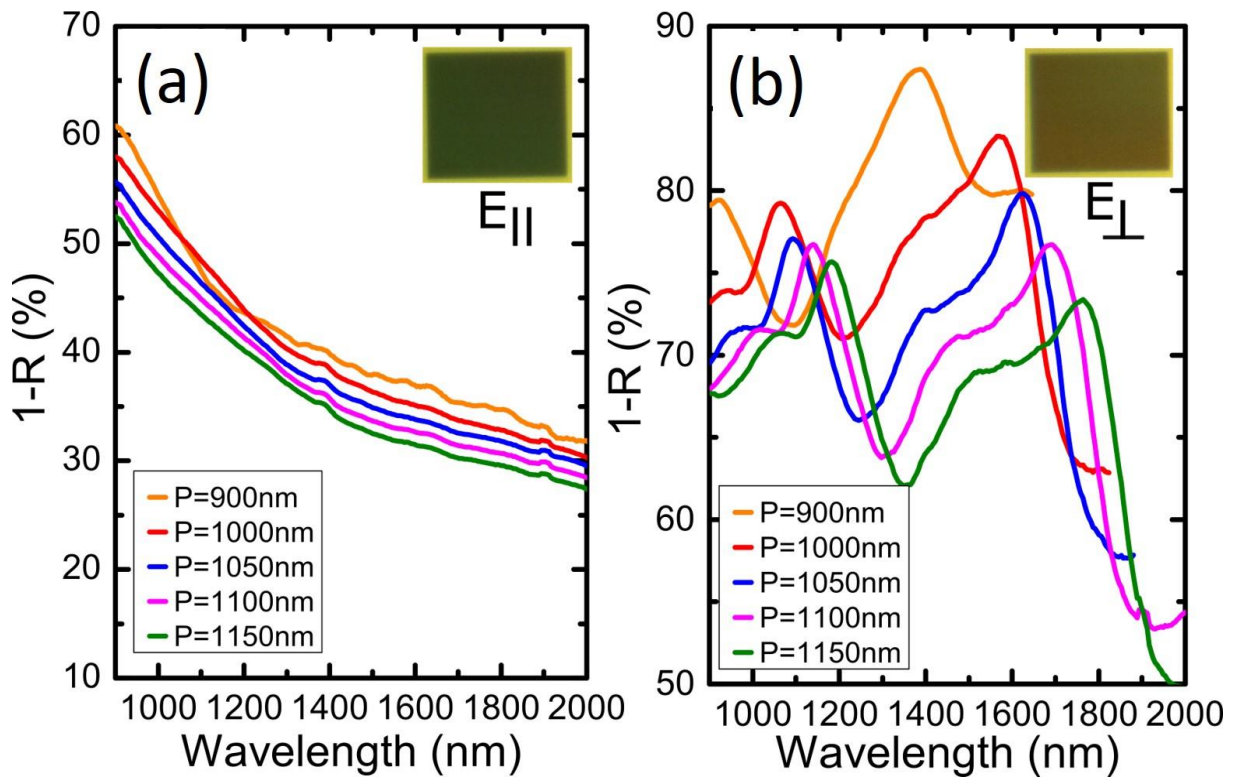

**Figure 3. Photo-response of the optical Nb nano-circuit arrays in the optical telecommunication frequencies:** Room temperature resonant response of various arrays with periods $P_y$ from 900 nm to 1150 nm for light polarized parallel (a) and perpendicular (b) to the nano-circuits, respectively. Plasmonic colours of the array with $P_y$= 900 nm is shown in the inset of each figure for different polarizations.

Several arrays of nano-gratings of 100x100 μm$^2$ area were milled by using a focused ion beam (FIB) into the Nb film surface, each with a groove length $L_x$=100 μm, width $w$=200 nm, and depth $d$=240 nm, and with grating periods varying between $P_y$=900 and 1150 nm. The high fabrication quality and uniformity of the sample are confirmed by scanning electron microscope as is shown by the SEM image in the inset of Fig. 2.

The room temperature resonant extinction 1-R of various grating arrays with periods $P_y$ from 900 to 1150 nm for light polarized parallel and perpendicular to the lines is shown in Fig. 3 (b) and (c) respectively. The arrays show plasmonic colours when illuminated using white light, as illustrated for grating array with $P_y$= 900 nm in the insets of Fig. 3 (b),(c) for the two different polarizations.

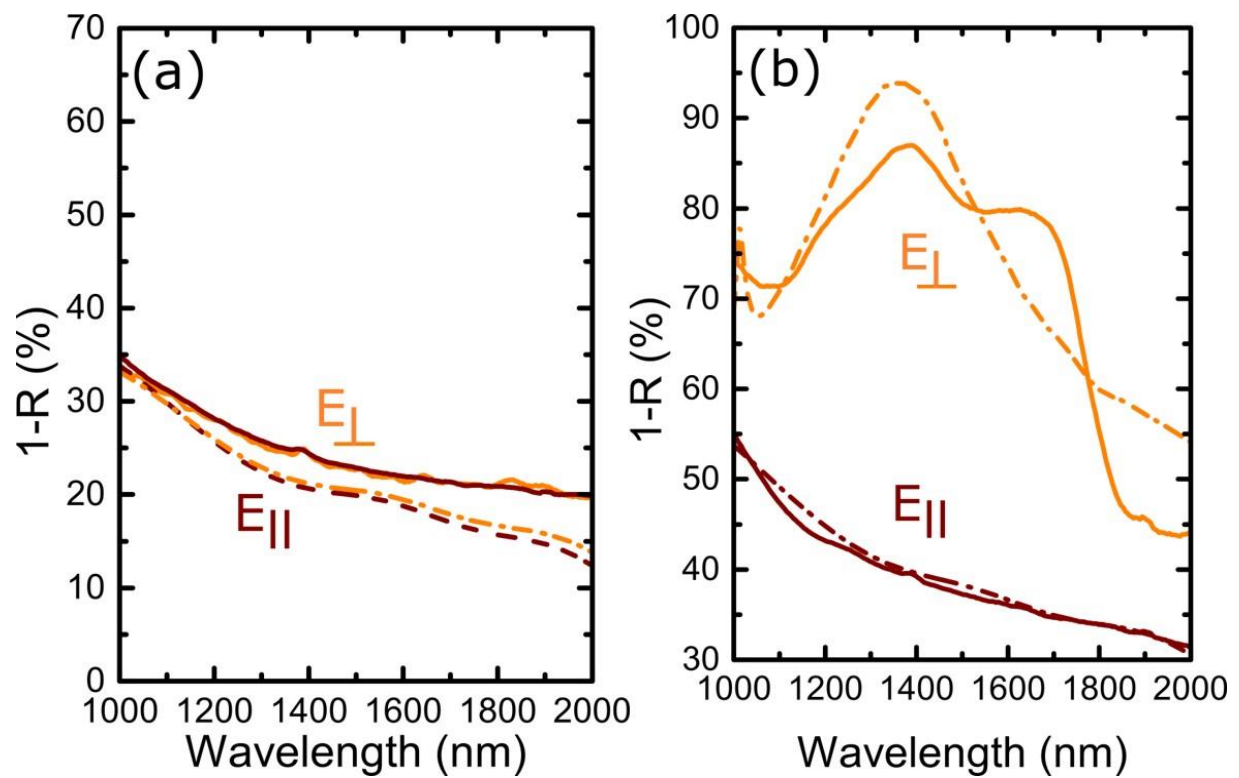

**Figure 4. Resonant response of unstructured Nb film and Nb grating array at Room Temperature.** (a) Absorption spectra of unstructured Nb film for vertical ($E_\perp$) and horizontal ($E_{II}$) polarizations at room temperature. (b) Absorption spectra of grating with $P_y$=900nm for two different polarizations. Solid lines: experiment, dashed-lines: Comsol modelling.

While the *x*-polarized response only shows a smooth variation of the reflectivity, resonance features are found for *y* polarization. Resonances in the extinction are observed for wavelengths longer than the grating period, indicating the functionality of the structure as a subwavelength metamaterial. The resonance is seen to shift to longer wavelengths with increasing grating period. Next to the prominent subwavelength resonance, a second feature consisting of adjacent dip and peak appears at slightly shorter wavelengths around the diffraction condition $P_y=\lambda$, corresponding to the Wood's and Rayleigh's anomalies of the 1st order grating diffraction.

In Fig. 4 (a) and (b) we compare the reflectivities of the unstructured Nb film and a representative Nb metamaterial with $P_y=$ 900nm, respectively, at room temperature and for two different illumination polarizations. Experimental data are confirmed by numerical simulations using COMSOL (dashed lines in Fig. 4a,b). Good agreement is obtained in the general behaviour for resonant and nonresonant responses, including the magnitude of the metamaterial resonance peak for *y*-polarization. Some deviation is seen for the precise shape of the resonance, in particular a shoulder in the extinction at around 1700 nm, which could be caused by variations in the fabrication process over the grating, resulting in inhomogeneous broadening and additional damping.

Figure 5 (a) and (b) shows the $E_y$- and $H_z$-filed distribution in the grating at the resonance wavelength of 1380 nm. The resonance shows a electrical dipole response in the field distribution in the groove.

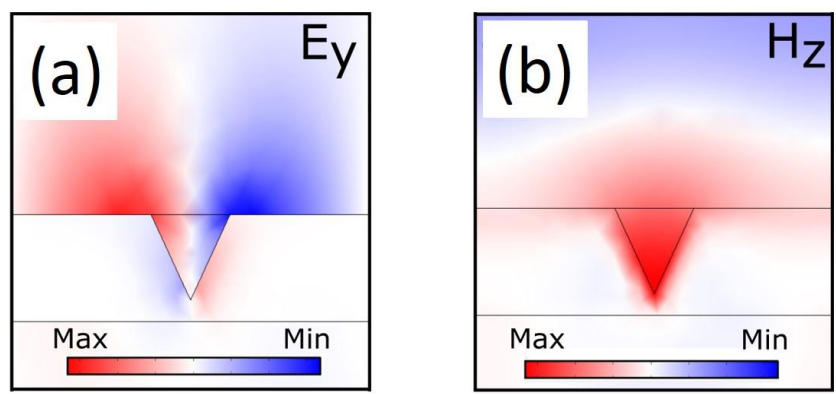

**Figure 5:** COMSOL modelling of the $E_y$ field (a), and $H_z$ field (b) distribution in the grating at the resonance wavelength of 1380 nm.

**Conclusion**

We demonstrated a new type of compact and chip-integrated optical nano-circuit arrays based on Nb. The device offer tunable plasmonic photo-response at fibre optic communication frequencies at room temperature. Given that Nb is a superconducting material with quantum mechanic phase below transition temperature ($T_c$ around 9 K), our results pave the way for the realisation of tunable photonic links, modulators, plasmonic single photon detectors, sensitive sensors and bolometers for applications in quantum computing and communication.